\documentstyle[sprocl]{article}

\input{psfig}

\def\Journal#1#2#3#4{{#1} {\bf #2}, #3 (#4)}

\def\be{\begin{equation}}
\def\ee{\end{equation}}
\def\bea{\begin{eqnarray}}
\def\eea{\end{eqnarray}}
\begin{document}

\title{POSITION EIGENSTATES AND THE STATISTICAL AXIOM OF QUANTUM MECHANICS}

\author{L. POLLEY}

\address{Physics Dept., Oldenburg University, 26111 Oldenburg,
Germany\\E-mail: polley@uni-oldenburg.de} 

\maketitle\abstracts{Quantum mechanics postulates the existence of 
states determined by a particle position at a single time. 
This very concept, in conjunction with superposition, induces much of the
quantum-mechanical structure. In particular, it implies the time evolution to
obey the Schr\"odinger equation, and it can be used to complete a truely basic 
derivation of the statistical axiom as recently proposed by Deutsch.}

\section{Quantum probabilities according to Deutsch}

A basic argument to see why quantum-mechanical probabilities must be squares 
of amplitudes (statistical axiom) was given by Deutsch \cite{Deutsch,BDeWitt}.
It is independent of the many-worlds interpretation. Deutsch considers a superposition
of the form
\begin{equation}             \label{A+B}
   \sqrt{\frac{m}{m+n}}~|A\rangle + \sqrt{\frac{n}{m+n}}~|B\rangle \qquad
   m,n\mbox{ integer}
\end{equation}
He introduces an auxilliary degree of freedom, $i=1,\ldots,m+n$, and 
replaces $|A\rangle$ and $|B\rangle$ by {\em normalized} superpositions,
\begin{equation}             \label{subst}
  |A\rangle  \longrightarrow  
                            \sqrt{\frac1m}\sum_{i=1}^m |A\rangle |i\rangle 
  \hspace{15mm}
  |B\rangle  \longrightarrow 
                            \sqrt{\frac1{n}}\sum_{i=m+1}^{m+n} |B\rangle |i\rangle 
\end{equation}
All amplitudes in the grand superposition are equal to $1/\sqrt{m+n}$ 
and should result in {\em equal} probabilities for the detection of 
the states. This immediately implies the ratio $m:n$ for the probabilities of 
property $A$ or $B$. 

The argument has clear advantages over previous derivations of the 
statistical axiom. Gleason's theorem \cite{Gleason,AsPeres}, 
for example, is mathematically non-trivial and not well received by many 
physicists, while von Neumann's assumption 
$\langle O_1 + O_2\rangle=\langle O_1 \rangle + \langle O_2 \rangle$
about expectations of observables \cite{Neumann,Ulfbeck}
is difficult to interpret physicswise if
$O_1$ and $O_2$ are non-commuting \cite{AsPeres,Neumann}. 
  
However, Deutsch's argument relies in an essential way on the unitarity of the 
replacement, 
or the normalization of any physical state vector. Why should a state vector 
be ``normalized'' in the usual sense of summing the squares of amplitudes? It would 
seem desirable to provide justification for this beyond its being ``natural'' 
\cite{BDeWitt}. In fact, the reasoning would appear circular without an extra 
argument about unitarity or normalization.  
I have proposed \cite{polley2} to realize the ``replacement'' (\ref{subst}) 
physically by the time evolution of a suitable device. Then, what can be said about 
quantum-mech\-an\-ic\-al evolution {\em without anticipating} the unitarity?

\section{Schr\"odinger's equation for a free particle as a consequence of 
         position eigenstates}

For free particles, a well-known and elegant way to obtain the Schr\"odinger 
equation is via unitary representations of space-time symmetries. Interactions can 
be introduced via the principle of local gauge invariance. However, this approach 
to the equation anticipates unitarity.

As I pointed out recently \cite{polley1}, the Schr\"odinger equation for a
free scalar particle is also a consequence of the very concept of a position 
eigenstate\footnote{Which relies on linear algebra, hence includes the 
concept of  ``superposition''.}
in discretized space. To an extent, this just means to regard ``hopping 
amplitudes'', as they are familiar from solid state theory, as {\it a priori} 
quantum-dynamical entities. The point is to show, however, that a 
hopping-parameter scenario {\em without} unitarity would lead to consequences 
sufficiently absurd to imply that unitarity must be a property of the physical 
system. As will be seen below, the absurdity is that a wave-function that makes 
perfect sense at $t=0$ would cease to exist anywhere in space at an earlier or later 
time. 

Consider a spinless particle ``hopping'' on a 1-dimensional chain of positions
$x=na$ where $n$ is integer and $a$ is the lattice spacing.    
\begin{center} \hspace*{15mm}
\unitlength=1.0cm
\begin{picture}(7.0,0.7)
%\put(-0.03,0.60){$\vdash$}
%\put( 0.43,0.65){$a$}
%\put( 0.82,0.60){$\dashv$}
\put(-0.03,-0.10){$\vdash$}
\put( 0.43,-0.05){$a$}
\put( 0.82,-0.10){$\dashv$}
%\put(2.92,0.8){\sf v}
%\put(3.5,0.8){\oval(1.0,1.0)[t]}
%\put(4.5,0.8){\oval(1.0,1.0)[t]}
%\put(4.92,0.8){\sf v}
\put(3.8,0.4){\vector(-1,0){0.6}}
\put(4.2,0.4){\vector( 1,0){0.6}}
%\put(3.43,0.6){$\beta$}
%\put(4.43,0.6){$\beta$}
\multiput(-1.0,0.4)(1.0,0.0){8}{\circle*{0.2}}
\put(2.76,0.0){$\scriptstyle n-1$}
\put(3.93,0.0){$\scriptstyle n  $}
\put(4.76,0.0){$\scriptstyle n+1$}
\end{picture}
\end{center}
Assume the particle is in an eigenstate $|n,t\rangle$ of position number $n$ at 
time $t$ (using the Heisenberg picture), and it has a possibility to change its
position. The information given by a ``position at one time'' does not
determine which direction the particle should go. Thus the eigenstate $|n,t\rangle$   
necessarily is a {\em superposition} when expressed in terms of eigenstates relating 
to another time $t'$. Moreover, because of the same lack of information, 
positions to the left and right will have to occur {\em symmetrically}.
If $t'\to t$, only nearest neighbours will be involved. Thus we expect
a ``hopping equation'' of the form   
$$ 
|n,t\rangle = \alpha~|n,t'\rangle + \beta~ |n+1,t'\rangle
                                  + \beta~ |n-1,t'\rangle
$$
This can be rewritten as a differential equation in $t$,
$$ 
-i\frac{{\rm d}}{{\rm d}t}~|n,t\rangle = V~|n,t\rangle + \kappa~ |n+1,t\rangle
 + \kappa~ |n-1,t\rangle \qquad \kappa, V \mbox{ complex (so far)}
$$ 
Parameters $\alpha,\beta$ and $\kappa,V$ are in an algebraic relation \cite{polley1}
which need not concern us here. To obtain an equation for a wave-function 
we consider a general state $|\psi\rangle$ composed of simultaneous position 
eigenstates,
$$ 
|\psi\rangle  = \sum_n \psi(n,t)~|n,t\rangle \qquad \mbox{(Heisenberg picture)}
$$
This defines the coefficients $\psi(n,t)$ for all $t$.
Now take the time derivative on both sides, identify $\psi(n,t)$ with 
a function $\psi(x,t)$ where $x=na$, and Taylor-expand the shifted values 
$\psi(x\pm a,t)$. This results in
$$ 
i\frac{\partial\psi}{\partial t} = 
(V+2\kappa)~\psi + a^2 \kappa \frac{\partial^2\psi}{\partial x^2} + {\cal O}(a^3)
$$
Finally, take $a\to 0$ on the relevant physical scale. The spatial spreading
of the wave-function is then given by the $a^2$ term, and the solution of the 
equation is
$$ 
    \psi(x,t) = e^{-i(V+2\kappa)t} \int \tilde{\psi}(p) 
                e^{i p x} e^{- i a^2\kappa p^2 t} {\rm d}p
$$
This time evolution would be unitary if $\kappa$ and $V$ were real. Hence, 
consider the consequences of a non-real $\kappa$. The integrand would then contain
an evolution factor increasing towards positive or negative times like   
$$
     \exp\left(\pm \, a^2 \, {\rm Im}\kappa \, p^2 \, t \right)
$$
This would lead to physically absurd conclusions about certain 
``harmless'' wave-functions, like the Lorentz-shape function $\psi(x)=1/1+x^2$:  
\begin{itemize}
\item For ${\rm Im}\,\kappa > 0$, ``harmless'' function
      $\tilde{\psi}(p) \propto \exp(-|p|)$ would not exist anywhere in space 
      after a short while.
\item For ${\rm Im}\,\kappa < 0$, the ``harmless'' function could not be prepared 
      for an experiment to be carried out on it after a short while.
\end{itemize}
In a mathematical sense, of course, it still remains a postulate that the value of
$\kappa$ be real. But physicswise, it does seem that unitarity of quantum mechanics
is unavoidable once the superposition principle and the concept of 
position eigenstate are taken for granted.

As for parameter $V$, the factor $e^{-iVt}$ would be raised to the $n$th 
power in an $n$-particle state, and would lead to an absurdity similar to the above
with certain superpositions of $n$-particle states unless $V$ is real,
too.

\section{Driven particle: Weyl equation in general space-time}

As an example of a particle interacting with external fields we may
consider a massless spin 1/2 particle with inhomogeneous hopping conditions
\cite{polley1}. Here the starting point is common eigenstates of spin and position, 
where ``position'' refers to a site on a cubic spatial lattice.
A particle in such a state at time $t$ will be in a superposition of neighbouring 
positions and flipped spins at a time $t'\approx t$. In 3 dimensions, and 
immediately in terms of a wave-function, the corresponding differential equation is 
$$ 
   -i \frac{\rm d}{{\rm d}t} \psi_s(\vec{x},t) =  
    \sum_{\rm lattice\atop directions} H_{nss'} \psi_{s'}(\vec{x}-a\hat{n},t)
$$
where $H_{nss'}$ are any complex amplitudes. On-site hopping  
(time-like direction) is included as $n=0$. 
To begin with, a free particle is defined by translational and rotational symmetry.
In this case, the hopping amplitudes reduce to two independent parameters 
\cite{polley1}, $\epsilon$ and $\kappa$, both of them complex so far. 
By Taylor-expanding the wave-function and taking $a\to0$ we find 
$$ 
   \partial_t \psi_s(\vec{x},t) = \epsilon \psi_s(\vec{x},t)
  - a \kappa\,\sigma^n_{ss'}  \partial_n \psi_{s'}(\vec{x},t)
$$
If $\kappa$ had an imaginary part, it would lead to physical absurdities with the 
time-evolution of certain ``harmless'' wave-functions similarly to the previous
section. For real $\kappa$, we recover the non-interacting Weyl equation. 

If we now admit for ``slight'' (order of $a$) anisotropies and inhomogeneities in 
the hopping amplitudes, by adding some $a\gamma_{\mu s s'}(\vec{x},t)$ to the 
hopping constants above, we recover a general-relativistic version of the equation 
\cite{Hehl} with the $\gamma_{\mu s s'}(\vec{x},t)$ acting as spin connection 
coefficients.  
Unitarity in this context means that the probability current density 
$$ 
     j^{\mu}(\vec{x},t) = \psi_{s}^*(\vec{x},t) ~ \sigma^{\mu}_{ss'}
                          \psi_{s'} (\vec{x},t)
$$
is covariantly conserved:
$$ 
    \partial_{\alpha} j^{\alpha} + \Gamma^{\alpha}_{\beta\alpha} j^{\beta} = 0
$$
This is found to hold automatically if the vector connection coefficients are 
identified as usual \cite{Hehl} through the matrix equation
$$ 
    \Gamma^{\mu}_{\nu\alpha} \, \sigma^{\nu} 
  + \sigma^{\mu} \gamma_{\alpha} + \gamma_{\alpha}^{\dag} \sigma^{\mu} = 0  
$$ 
Imposing no constraints on the spin connection coefficients,
we are dealing with a metric-affine space-time here, which can have torsion and
whose metric may be covariantly non-constant. The study of space-times of this 
general structure has been motivated by problems of quantum gravity \cite{Hehl}.
It may be interesting to note that nothing but 
{\em propagation by superposing next-neighbour states} needs to be assumed here.
In particular, scalar products of state vectors are not needed.

\section{Realizing Deutsch's ``substitution'' as a time evolution}

\begin{figure}[t]
%\rule{5cm}{0.2mm}\hfill\rule{5cm}{0.2mm}
%\vskip 2.5cm
%\rule{5cm}{0.2mm}\hfill\rule{5cm}{0.2mm}
\hfill \psfig{figure=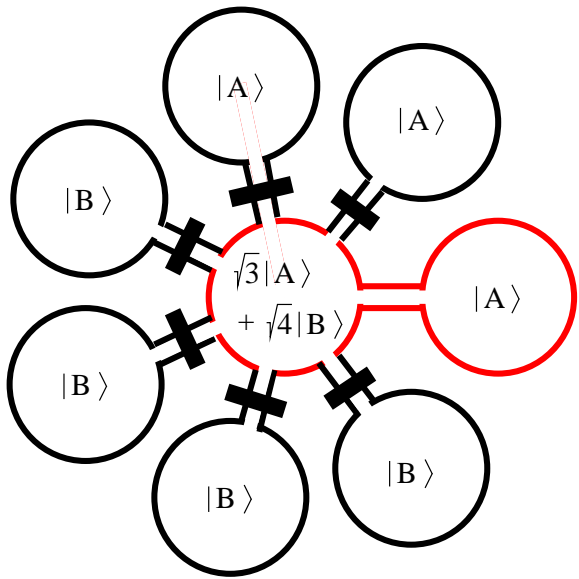,height=3.0in} \hfill~
\caption{An array of eight cavities of equal shape.
The initial state is located in the central cavity.
When each channel is opened for an appropriate time,
the state evolves to an equal-amplitude superposition
of the peripheral cavity-states. \label{fig:cavities}}
\end{figure}
Having demonstrated ``automatic'' unitarity on two rather general examples
we can now turn with some confidence to the original issue of completing Deutsch's
derivation of the statistical axiom.  

To realize the particular substitution (\ref{subst}) for state vector 
(\ref{A+B}), let us consider a particle with internal eigenstate $|A\rangle$ 
or $|B\rangle$, such as the polarisations of a photon. Let this particle be placed 
in a system of cavities\footnote{Or Paul traps, or any other sort of potential well; 
these are to enable us to store away parts of the wave function so that there is 
no influence on them by the other parts.} connected by channels 
(Fig.\,\ref{fig:cavities}) which can be opened selectively for internal state 
$|A\rangle$ or $|B\rangle$. It will be essential
in the following that all cavities are of the same shape, because this will enable
us to exploit symmetries to a large extent. The location of the 
particle in a cavity will serve as the auxilliary degree of freedom as in 
(\ref{subst}), except that $|A\rangle$ and $|B\rangle$ before the substitution will
be identified with $|A\rangle|0\rangle$ and $|B\rangle|0\rangle$ where $|0\rangle$ 
corresponds to the central cavity. 

Now let only one of the channels be open at a time. We are then dealing
with the wave-function dynamics of a two-cavity subsystem, while the rest of 
the wave-function is standing by. What law of evolution could we expect? 
A particle with a well-defined (observed) position $0$ at time $t$ will no longer 
have a well-defined position at time $t'$ if we allow it to pass through a channel,
without observing it. Thus a state $|0,t\rangle$ defined by position $0$ at time $t$ 
(using the Heisenberg picture) will be a superposition when expressed in terms of 
position states relating to a different time $t'$. In particular, if channel 
$0\leftrightarrow1$ is the open one, 
$$ 
    |0,t\rangle = \alpha |0,t'\rangle + \beta  |1,t'\rangle 
$$
Likewise, by symmetry of arrangement,  
$$
    |1,t\rangle  = \alpha |1,t'\rangle  + \beta  |0,t'\rangle 
$$
It follows that $|0,t\rangle\pm|1,t\rangle$ are stationary states  
whose dependence on time consists in prefactors
\begin{equation}            \label{a+-b}
    (\alpha \pm \beta)^k  \qquad \mbox{after $k$ time steps.}
\end{equation}
If the particle is initially in the rest of the cavities, whose channels are shut, 
we would expect this state not to change with time:  
$$
    |{\rm rest},t\rangle = |{\rm rest},t'\rangle
$$
Now, if (\ref{a+-b}) were {\em not} mere phase factors, we could easily construct 
a superposition of $|0\rangle$, $|1\rangle$, and $|{\rm rest}\rangle$ so that,
{\em relative to the disconnected cavities}, the part of the state vector 
in the connected cavities would grow indefinitely or vanish in the long run. 
As there is no physical reason for such an imbalance between the connected and the 
disconnected cavities, we conclude that 
$$ 
    \alpha + \beta = e^{i\varphi}   \qquad  \qquad 
    \alpha - \beta = e^{i\varphi'}
$$ 
Having shown evolution through one open channel to be unitary, we can identify an 
opening time interval \cite{polley2}, $\tau_m$, to realize the following step of 
the replacement (\ref{subst}):
$$
    \sqrt{m}|A\rangle |0\rangle + |{\rm rest}\rangle 
     \quad \stackrel{\tau_m}{\longrightarrow}\quad
      \sqrt{m-1}|A\rangle |0\rangle + |A\rangle |1\rangle + |{\rm rest}\rangle
$$
Here $|{\rm rest}\rangle$ stands for state vectors that are decoupled,
such as all $|B\rangle |i\rangle$, and all $|A\rangle |i\rangle$ with $i\neq0,1$.
Opening other channels analogously, each one for the appropriate $\tau_m$ and 
internal state, we produce an equal-amplitude superposition 
$$
    \sum_{i=1}^m |A\rangle |i\rangle + \sum_{i=m+1}^{m+n} |B\rangle |i\rangle
$$
The probability of finding the particle in a particular cavity is now $1/m+n$
as a matter of symmetry. As the internal state is correlated with a cavity by
the conduction of the process, the probabilities for $A$ and $B$ immediately 
follow. These must also be the probabilities for finding $A$ or $B$ in the 
original state, because properties $A$ and $B$ have remained unchanged during the 
time evolution.

\section{Can normalization be replaced by symmetry?}

An interesting side effect of the above realization of Deutsch's argument is that 
state vectors need no longer be normalized at all. Permutational symmetry of a
superposition suffices to show that all possible outcomes of an experiment must 
occur with {\em equal}\/ frequency. Then the numerical values of the probabilities 
are fully determined.
This feature of quantum probabilities may be relevant to problems of normalization 
in quantum gravity \cite{Asht}, such as the non-locality of summing $|\psi|^2$ over 
all of space, or the non-normalizability of the solutions of the Wheeler-DeWitt 
equation.

\end{document}